# Connecting science fundamentals and career proficiency

# &

# Development of multi-disciplinary science curriculum

F.A. Selim[*]

Department of Physics and Astronomy, Bowling Green State University, OH 43403

[*]faselim@bgsu.edu

**Abstract**

The current conventional approach in teaching STEM seems inadequate; it is failing both the students and the work-force demands in several aspects. Perhaps it is succeeding in equipping students with information and facts but surely falls short in developing them intellectually and in sparking their passion for science. Mixing research with learning by adding components in courses to enforce creativity, reasoning, and to connect the dots between different topics and different subjects can be an effective counter to this tendency. Multi-disciplinary science curriculum connecting the dots between different subjects and the dots between fundamentals and applications and demonstrating clear path to current and future industrial jobs can offer a good approach for STEM education. An example for mutli-disciplinary science course is presented.

**Introduction**

Dramatic advances in science and technology in recent decades are now inextricably integrated into all aspects of modern life from health care to manufacturing, energy, and communications. And yet integration in the educational realm has not kept similar pace (Atkin 2005). Although awareness of the need to reform STEM education is widespread (Tobias 1992, Wicklein 1995), science courses are still generally taught in a centuries-old model of separate, non-overlapping disciplines (Hake 2007). This contributes to the high attrition rate in STEM fields as the abstract nature of introductory courses frequently fails to capture students' imagination. Students appear to have trouble 'connecting the dots', linking the foundational classes they experience with the prospective careers to which they aspire. Moreover, STEM fields are the major driving force in job creation and economic growth and workforce development needs are changing rapidly (Lewis 2014). For instance, in health care the new area of nanomedicine builds on technological innovation of nanodevices combined with computer-controlled procedures for both diagnosis and treatment. Progress in diverse sectors of industry and academia thus depend crucially on individuals with expertise and strong multidisciplinary training in physics, chemistry, mathematics, and biology.

Innovative experimental courses targeting undergraduate science majors integrate basic physics, biological-, and chemical principles are needed to demonstrate how interdisciplinary thinking generates real-world problem solving. For instance, one might connect theory to application by employing widely used spectroscopy/microscopy techniques as a vehicle to teach advanced chemical- and quantum mechanical principles in an accessible and meaningful way. Developing multi-disciplinary science curriculum connecting fundamental sciences to applications, and crossing the boundaries between physics, chemistry, mathematics, and biology can make theories more accessible to the students.

**Development of multi-disciplinary science course**

The need for disciplinary approaches to integrate science has been widely recognized (Fairweather 2008, Kali 2008), yet no significant progress has been made in universities and colleges across the nation, and the question remains, how to implement new interdisciplinary approaches in the current curriculum in meaningful ways. As a good starting point based on physics but connected to other subjects, we can develop a new multi-disciplinary course to address specific real-world problems through the integration of quantum mechanics with physical, biological and chemical principles. The course should be taught outside the traditional boundaries of physics, geology, chemistry, and biology and instead apply their basic principles in interdisciplinary topics such as medicine, pharmaceuticals, technology, clean energy resources, food, environment, and land resources. A large proportion of current and future STEM jobs are associated with these topics in some form. Therefore, preparing students with a strong diverse background in the basic sciences and equipping them with hands-on experience of technologies directly applied to these topics would boost both their personal development and economic growth in these sectors. I propose a course titled "Integrated techniques for Multidisciplinary Science" and focused on most advanced techniques commonly applied in the physical and life sciences such as:

1. Imaging and microscopy techniques (X-ray imaging, X-ray crystallography, scanning electron microscopy, transmission electron microscopy, atomic force microscopy): It is well recognized that such techniques are crucial in all areas of research, industry, diagnostics, drug development, security and more. They are based on quantum mechanics theories (wave particle duality, electron diffraction, etc) that are ambiguous to the majority of science students and graduates despite the widespread use of these techniques, even physics majors experience difficulties in picturing and grasping quantum mechanical principles (Singh 2006). The proposed approach - teaching the theory intimately connected to techniques and applications - would facilitate students' persistence in grappling with such difficult concepts. Through studying the basic science behind these techniques and applying them to real problems, the students can ultimately develop a much deeper understanding of particle-wave duality and diffraction theories.

2. Nuclear magnetic resonance (NMR) and electron spin resonance (ESR) techniques: The students may learn the basic atomic physics connected with the nuclear and electron spin and magnetic moments and recognize how these have been harnessed to create powerful tools for medical diagnostics, as well as the structural determination of biomolecular molecules, which is crucial in drug development and materials research.

3. Optical and vibrational spectroscopies (Ultraviolet and visible absorption (UV-VIS) spectroscopy, reflectance, Raman and Fourier transform spectroscopies): Through teaching these techniques, one can introduce the students to electromagnetic waves and quantum chemistry, their applications in the chemical identification of elements, the study of biological- and chemical molecules, and geological samples.

4.Widespread luminescence techniques such as fluorescence and chemo-luminescence. These are widely used in biology, geology, and chemistry laboratories, as well as in medical diagnostics. The students will learn about the quantum theory of light and interaction of light with matter.

These topics are selected as they are commonly used in a wide range of physical and life science applications and are not limited to one or two fields. In each of these topics, the students should learn the basic sciences behind the technique, use it in laboratories and apply it to actively investigate a real problem in the area which appeals most to their interests - food, pharmaceuticals, materials science, technology, or clean energy, to name just a few.

**Course delivery**: The course should be offered to all science majors who have completed the introductory physics and chemistry classes. It can be taught in three modes of delivery. First, interactive lectures with a conceptual focus, eliminate the confounding details to provide a solid understanding of the basic science underlying each technique. Second, in accompanying laboratory sessions, students participate in hands-on technical training and acquire skills in data analysis. Working in laboratories to solve a real world-problem can provide students opportunities to develop their critical thinking abilities and become more skilled in identifying effective solutions. Finally, peer instruction in the form of proposal preparation sessions where students are assigned individual projects for which they must determine appropriate techniques and draft a research plan. A range of assessment tools should be used to both guide and gauge performance on coursework. These may include traditional summative (e.g. quizzes, exams) and formative assessments (e.g. peer assessments, presentations) to measure the sophistication of the student's technical and analytical knowledge.

**Conclusion**

the proposed multi-disciplinary approach and strong connection between fundamental concepts and applied sciences in the proposed course would improve student retention in STEM field. According to the National Science Foundation report (National Science Foundation; Center for Science Mathematics and Engineering Education, Committee on Undergraduate Science Education 1999) teaching practices in college STEM courses appear to be behind the decline number of students in STEM field. The proposed integrated approach should help to recruit and retain more students in STEM fields, enhance their understanding of science culture, and provide them tools which are immediately useful for their careers in high-tech industries or research. Clearly one course is insufficient to cover the advanced techniques associated with all aspects of modern life and technologies. The proposed course could be an example and step towards development of a comprehensive multi-disciplinary STEM degree program.

- Wicklein, R. C., & Schell, J. W. (1995). Case studies of multidisciplinary approaches to integrating mathematics, science, & technology education.